\documentclass[aps,amsmath,amssymb,twocolumn]{revtex4}

\usepackage{graphicx}
\usepackage{dcolumn}
\usepackage{bm}
\usepackage{amsmath}
\usepackage{epstopdf}
\usepackage{amsfonts}
\usepackage{amssymb}
\usepackage{epsfig}
\usepackage{tabularx}
\usepackage{color}
\usepackage{soul}

\usepackage[colorlinks=true,citecolor=blue,urlcolor=magenta,breaklinks]{hyperref}

\newcommand{\be}{\begin{equation}}
\newcommand{\en}{\end{equation}}
\newcommand{\bea}{\begin{eqnarray}}
\newcommand{\ena}{\end{eqnarray}}

\begin{document}

\title{Orbits of light rays in scale-dependent gravity: Exact analytical solutions to the null geodesic equations}

\author{
Grigoris Panotopoulos  {${}^{a}$
\footnote{
\href{mailto:grigorios.panotopoulos@tecnico.ulisboa.pt}{grigorios.panotopoulos@tecnico.ulisboa.pt} 
}
}
\'Angel Rinc\'on {${}^{b}$
\footnote{
\href{mailto:aerinconr@academicos.uta.cl}{aerinconr@academicos.uta.cl} 
}
} 
Il{\'i}dio Lopes  {${}^{a}$
\footnote{
\href{mailto:ilidio.lopes@tecnico.ulisboa.pt}{ilidio.lopes@tecnico.ulisboa.pt} 
}
}
}

\address{
${}^a$ Centro de Astrof{\'i}sica e Gravita{\c c}{\~a}o-CENTRA, Instituto Superior T{\'e}cnico-IST, Universidade de Lisboa-UL, Av. Rovisco Pais, 1049-001 Lisboa, Portugal.   
\\
${}^b$ Sede Esmeralda, Universidad de Tarapac\'a,
Avda. Luis Emilio Recabarren 2477, Iquique, Chile.
}

\begin{abstract}
We study photon orbits in the background of $(1+3)$-dimensional static, spherically symmetric geometries. 
In particular, we have obtained exact analytical solutions to the null geodesic equations for light rays 
in terms of the Weierstra{\ss} function for space-times arising in the context of scale-dependent gravity. 
The trajectories in the $(x-y)$ plane are shown graphically, and we make a comparison with similar geometries 
arising in different contexts. The light deflection angle is computed as a function of the running parameter $\xi$,
and an upper bound for the latter is obtained.
\end{abstract}

\maketitle

%%%%%%%%%%%%%%%%%%%%%%%
\section{Introduction}
%%%%%%%%%%%%%%%%%%%%%%

Light has always been of paramount importance in the history of science. Indeed, over the years considerable
progress has been made observing the light reaching our planet from distant sources. To mention just a few,
the absorption spectra of chemical elements, the accidental discovery of the cosmic microwave background radiation 
by Penzias and Wilson \cite{PenWil}, and the bending of light during the total solar eclipse in 1919, are only some 
examples among many others. As far as gravity is concerned, studying the motion of light rays and/or massive 
test particles in a fixed gravitational background is one of the principal ways to explore the physics of a 
given gravitational field. For instance, in the case of Einstein's General Relativity (GR) \cite{Einstein:1916vd} 
and Schwarzschild geometry \cite{Schwarzschild:1916uq}, the explanation of the perihelion precession of the planet 
Mercury around the Sun \cite{deSitter:1916zza} as well as the bending of light \cite{deSitter:1916zza} 
during the total solar eclipse in May of 1919 (for a historical review for the completion of 100 years 
of that important event and the two British expeditions to Sobral, Brazil, and to Pr{\'i}ncipe Island, 
Africa, see \cite{Crispino:2019yew}) comprise two of the classical tests of GR \cite{tests}.

\smallskip 

Moreover, understanding how light propagates through space in the presence of massive bodies is critical to our understanding of the Universe, e.g. to characterize the nature of dark energy and dark matter. Indeed, to understand the gravitational lensing of distant galaxies, we need to know precisely how light bends near the strong gravitational fields 
of galaxies. This is a central piece of any cosmological model \cite{M1}, such as the standard $\Lambda$CDM model \cite{M2}.

\smallskip

Despite its beauty and its many successful predictions that have been confirmed over the years, 
there are still numerous open questions concerning the quantum nature of GR.
The quest for a theory of gravity that incorporates quantum mechanics in a consistent way is still one of the major 
challenges in modern theoretical physics. Most current approaches to the problem found in the literature (for a 
partial list see e.g. \cite{QG1,QG2,QG3,QG4,QG5,QG6,QG7,QG8,QG9} and references therein), seem to share one 
particular property, i.e. the couplings that enter into the action defining one's favorite model, such as the 
cosmological constant, the gravitational and electromagnetic couplings etc, become scale dependent (SD) quantities 
at the level of an effective averaged action after incorporating quantum effects. 
A posteriori this does not come as a surprise, since scale dependence at the level of the effective action is a generic  feature of ordinary quantum field theory.

\smallskip
 
Due to the complexity of non-linear partial differential equations, analytical methods in most of the cases
just cannot work, and most of the gravitational effects can only be understood employing either approximate 
or numerical methods. Obtaining exact analytic expressions, however, is always desirable for at least two reasons. 
The first reason is that analytic expressions may serve as test beds for numerical methods, and they are also a 
good starting point for developing approximate approaches \cite{evaAIP}. The second reason is that analytic expressions 
enable a systematic study of the complete parameter space, and of 
all possible solutions, the structure and characteristics of which can be explored \cite{evaAIP}. This 
includes the derivation of observable effects, such as the perihelion advance or light deflection. It is only in 
this case that a systematic study of all effects may be performed. 

\smallskip

The orbits of light rays in fixed gravitational backgrounds of certain forms are described by
solutions to differential equations of elliptic or hyperelliptic type. The theory of those functions was 
studied long time ago by Jacobi \cite{Jacobi}, Abel \cite{Abel}, Riemann \cite{Riemann1,Riemann2} and 
Weierstra{\ss} \cite{Weierstrass}. A review of those achievements as well as a compact description of the 
complete theory may be found in \cite{Baker}. In particular, the motion of test particles in the 
Schwarzschild space-time was completely analyzed by 
Hagihara using elliptic functions back in 1931 \cite{Hagihara}. More recently, over the last 15 years or 
so, elliptic and hyperelliptic functions have been used to obtain exact analytic solutions to the geodesic 
equation for well-known geometries, such as Schwarzschild-(anti) de Sitter 
\cite{Kraniotis:2003ig,Hackmann:2008zza,Hackmann:2008zz}, Kerr \cite{Kerr:1963ud} with a non-vanishing 
cosmological constant \cite{Kraniotis:2004cz,Hackmann:2010zz}, regular black holes 
\cite{AyonBeato:1998ub,AyonBeato:1999rg,Balart:2014cga,Garcia:2013zud}, and higher-dimensional 
space-times \cite{Hackmann:2008tu,RN5D}.

\smallskip

Over the last years scale-dependent gravity has emerged as an interesting framework with appealing properties. 
As it is inspired by the renormalization group approach, it naturally allows for a varying cosmological constant 
and a varying Newton's constant at the same time. Its impact on black hole physics has been investigated in detail 
%\cite{previous,Koch:2014joa,angel,Rincon:2017ypd,Rincon:2017ayr,Hernandez-Arboleda:2018qdo,Contreras:2017eza,Contreras:2018dhs,Contreras:2018swc,Rincon:2018lyd,Rincon:2021hjj,Panotopoulos:2021obe,Panotopoulos:2020mii,Rincon:2020cpz,Panotopoulos:2020zqa}.
\cite{SD1,SD0,SD2,SD3,SD4,SD5,SD6,SD7,SD8,SD9,SD10,SD11,SD12,SD13,SD14,SD15,SD16}, 
while recently some astrophysical and cosmological implications were
studied as well \cite{astro1,astro2,cosmo1,cosmo2}. In the present work we propose to study the orbits of light 
rays in the background of two four-dimensional static, spherically symmetric space-times in the framework of scale-dependent gravity. Our work in this article is organized as follows: After this introduction, in the next section we briefly review 
the background geometry as well as the equations of motion for test particles. In section three we focus on 
null geodesics for light rays, and obtain exact analytical solutions describing photon orbits in a fixed gravitational
field for geometries arising in several different contexts other than GR . Finally, we close 
our work in the last section with some concluding remarks.

%%%%%%%%%%%%%%%%%%%%%%%%%%%%%%%%%%%%%%%%%%%%%%%%%%%%%%%%%%%%%%%%%%%%%%
\section{Background geometry and equations of motion for test particles}
%%%%%%%%%%%%%%%%%%%%%%%%%%%%%%%%%%%%%%%%%%%%%%%%%%%%%%%%%%%%%%%%%%%%%%

\subsection{Background geometry}

Scale-dependent gravity is a formalism able to extend classical GR solutions via the inclusion of scale-dependent couplings which account for quantum features. Notice that the corrections are assumed to be small. Avoiding the details, in absence of matter content, we have two "running" couplings: i) Newton's constant $G_k$ and ii) the cosmological constant $\Lambda_k$. We can also define the parameter $\kappa_k \equiv 8 \pi G_{k}$, with $G_k$ being the running Newton's constant, alternatively as the Einstein's constant. In addition, two extra fields need to be included: i) the metric tensor $g_{\mu \nu}$ and ii) the arbitrary renormalization scale $k$.
The effective Einstein's field equations, considering scale-dependent couplings, are given by \cite{SD8,SD10}
\begin{equation}
G_{\mu \nu } + \Lambda_k g_{\mu \nu} \equiv \kappa_k T_{\mu \nu}^{\text{effec}},
\end{equation}
with $\Lambda_k$ being the running cosmological constant, where the effective energy-momentum tensor is defined by  \cite{SD8,SD10}
\begin{equation}
\kappa_k T_{\mu \nu}^{\text{effec}} =  \kappa_k T_{\mu \nu}^{M} - \Delta t_{\mu \nu}.
\end{equation}
where $T_{\mu \nu}^{M}$ is the stress-energy tensor of matter (if any), while the additional contribution due to 
the scale-dependent Newton's constant is computed to be \cite{SD8,SD10}
\begin{equation}
\Delta t_{\mu \nu} \equiv G_k \Bigl( g_{\mu \nu} \square - \nabla_{\mu} \nabla_{\nu} \Bigl) G_k^{-1}. 
\end{equation}
To close the system of equations, we take advantage of the null energy condition, we accept 
that $\mathcal{O}(k(r)) \rightarrow \mathcal{O}(r)$, and we only consider the radial variable.

Finally, the scale-dependent Schwarzschild-de Sitter (SdS) solution is found to be \cite{SD8,SD10}
\begin{align}
\begin{split}
f(r)  =  f_{\text{SdS}}(r) + & \frac{\xi}{2M}  
\Bigg[
6 M - 2r + \frac{3 r \xi}{M}  (r - 4  M)
\\
& +
 \frac{2 r^2 \xi}{M}  (1 + 6 \xi) \ln \left( 1 + \frac{M}{r \xi } \right)
\Bigg]
\end{split}
\end{align}
where $f_{\text{SdS}}(r)$ is the lapse function of the usual SdS space-time given by
\begin{align}
f_{\text{SdS}}(r) \equiv 1 - \frac{2 M}{r} -\frac{1}{3} \Lambda  r^2
\end{align}
with $M$ being the mass of the object that generates the gravitational field, $\Lambda$ is the classical cosmological constant, and $\xi$ is a dimensionless parameter that measures the deviation from the classical theory.

Assuming that $\xi \ll 1$, taking the leading terms in $\xi$ we obtain the approximate expression \cite{SD10}
\begin{align}
f(r) \approx & \ f_{\text{SdS}}(r) -   \frac{\xi r}{M}   \left(1 - \frac{3M}{r}\right) \: 
\end{align}
Therefore in the following we shall consider static, spherically symmetric geometries of the form
\begin{equation} \label{background}
ds^2 = -f(r) dt^2 + f(r)^{-1} dr^2 + r^2 [ d \theta^2 + \sin^2\theta d \phi^2 ]
\end{equation}
characterized by a lapse function $f(r)$ of the general form
\begin{align}
f(r) & = f_{\text{SdS}}(r) + \gamma r + \eta
\end{align}
where the parameters $\gamma, \eta$ may be either positive or negative. Interestingly enough, it turns out that this is a class of geometries including solutions in contexts other than GR and scale-dependent gravity, such as Weyl conformal gravity \cite{Mannheim:1988dj} and massive gravity \cite{deRham:2010kj,Panpanich:2018cxo}. For comparison reasons, in the discussion to follow we shall show the photon orbits in the $(x-y)$ plane for all three space-times.

Here we should keep in mind that a regular scale-dependent black hole solution was first obtained in \cite{Contreras:2017eza} (see also \cite{Sendra:2018vux}), and the corresponding lapse function is found to be
\begin{eqnarray}
F(r)=1-\frac{2 M}{r} \left( 1+\frac{M \xi }{6 r} \right)^{-3},
\end{eqnarray}
where $M$ is the mass of the black hole, and $\xi$ is the scale-dependent parameter, which measures the deviation from the classical theory. Expanding in powers of $\xi$ we obtain an approximate expression at leading order in $\xi$
\begin{align}
F(r) & = 1 - \frac{2 M}{r} + \frac{M^2 \xi}{r^2} + \mathcal{O}(\xi^2)
\end{align}
Notice that at this level of approximation, the geometry looks like the Reissner-Nordstr{\"o}m solution \cite{RN} for charged black holes in GR, where $\xi$ plays the role of the electric charge. The orbits of test particles in the 
background of RN may be found precisely as in the Schwarzschild case \cite{oldpaper}. In the present work, for comparison reasons we shall show the photon orbits for this geometry as well.

\subsection{Geodesic equations}

As already mentioned before, we assume a fixed four-dimensional static, spherically symmetric gravitational background, which takes the general form
\begin{equation}
ds^2 = g_{tt} dt^2 - g_{rr} dr^2 - r^2 [ d \theta^2 + \sin^2\theta d \phi^2 ]
\end{equation}
The equation of motion for test particles is given by the geodesic equation \cite{Garcia:2013zud}
\begin{equation}
\frac{d^2x^\mu}{ds^2} + \Gamma^\mu_{\rho \sigma} \frac{dx^\rho}{ds} \frac{dx^\sigma}{ds} = 0
\end{equation}
where $s$ is the proper time, while the Christoffel symbols $\Gamma^\mu_{\rho \sigma}$ are computed by \cite{landau}
\begin{equation}
\Gamma^\mu_{\rho \sigma} = \frac{1}{2} g^{\mu \lambda} \left( \frac{\partial g_{\lambda \rho}}{\partial x^\sigma} + \frac{\partial g_{\lambda \sigma}}{\partial x^\rho} - \frac{\partial g_{\rho \sigma}}{\partial x^\lambda} \right)
\end{equation}
The mathematical treatment is considerably simplified by the observation that there are two first integrals of motion (i.e., conserved quantities), precisely as in the Keplerian problem in classical mechanics.
To do that, we recognize that for $\mu=1=t$ and $\mu=4=\phi$ the geodesic equations take the simple form
\begin{eqnarray}
0 & = & \frac{d}{ds} \left( g_{tt} \frac{dt}{ds} \right) \\
0 & = & \frac{d}{ds} \left( r^2 \frac{d\phi}{ds} \right) 
\end{eqnarray}
Taking advantage of the last two expressions, we can introduce the following conserved quantities
\begin{equation}
E \equiv g_{tt} \frac{dt}{ds}, \; \; \; \; \; \; L \equiv r^2 \frac{d\phi}{ds}
\end{equation}
The above two quantities, $\{E, L\}$, are identified to the energy and angular momentum, respectively. Moreover, assuming a motion on the $(x-y)$ plane the geodesic equation corresponding to the $\theta$ index is automatically satisfied.

Thus, the only non-trivial equation is the one corresponding to $\mu=2=r$ \cite{Garcia:2013zud}
\begin{equation}
\left( \frac{dr}{ds} \right)^2 = \frac{1}{g_{tt} g_{rr}} \: \left[ E^2 - g_{tt} \left( \epsilon + \frac{L^2}{r^2} \right) \right]
\end{equation}
which may be also obtained from \cite{Garcia:2013zud}
\begin{equation}
g_{\mu \nu} \frac{dx^\mu}{ds} \frac{dx^\nu}{ds} = \epsilon
\end{equation}
where $\epsilon = 1$ for massive test particles, and $\epsilon = 0$ for light rays.

At this point is advantageous to introduce an effective potential
\begin{align}
V^2_{\text{eff}} &= g_{tt} \left( \epsilon + \frac{L^2}{r^2} \right)
\end{align}
after which the equation of motion takes the final simple form \cite{Garcia:2013zud}
\begin{equation}
\left( \frac{dr}{ds} \right)^2 = \frac{1}{g_{tt} g_{rr}} \: \left[ E^2 - V^2_{\text{eff}} \right]
\end{equation}
Finally, the orbit is found obtaining $r$ as a function of $\phi$, which is found to be
\begin{equation}
\left( \frac{dr}{d\phi} \right)^2 = \left( \frac{dr/ds}{d\phi / ds} \right)^2  = \frac{r^4}{L^2 \: g_{tt} g_{rr}} \: \left[ E^2 - g_{tt} \left( \epsilon + \frac{L^2}{r^2} \right) \right]
\end{equation}
When the Schwarzschild ansatz is used (as in all cases of interest here), the latter expression is simplified to be
\begin{equation} \label{Arr}
R(r) \equiv \left( \frac{dr}{d\phi} \right)^2 = \frac{r^4}{L^2} \: \left[ E^2 - g_{tt} \left( \epsilon + \frac{L^2}{r^2} \right) \right]
\end{equation}
The differential equation, since it is of first order, must be supplemented by the appropriate initial condition. 
Clearly, the precise shape of the orbits depends on i) the background geometry, ii) the values of $(E,L)$, and iii) the initial condition.

%%%%%%%%%%%%%%%%%%%%%%%%%%%%%%%%%%
\section{Exact analytic solutions}
%%%%%%%%%%%%%%%%%%%%%%%%%%%%%%%%%%%

In this section, first we shall obtain the general expression for the orbits in terms of the Weierstra{\ss} function \cite{Weierstrass}, and after that we shall show graphically the orbits in the $(x-y)$ plane for different values of the 
photon energy and different initial conditions, and for each geometry separately.

\subsubsection{Case I}

The general case take advantage of the metric components as follows
\begin{align} \label{metric1}
g_{tt} &=f(r) = 1 - \frac{2 M}{r} - \frac{1}{3} \Lambda  r^2
+ \gamma  r + \eta \\
g_{rr} &= f(r)^{-1}
\end{align}
We recall that the above lapse function arises in several non-standard theories of gravity, such as Weyl gravity \cite{Mannheim:1988dj}, massive gravity \cite{deRham:2010kj,Panpanich:2018cxo}, and scale-dependent gravity  \cite{SD1,SD0,SD8,SD10} 
at leading order in $\xi$.

Focusing on photon orbits we set $\epsilon=0$, and making the change of variable $u=1/r$, the equation for the trajectories $u(\phi)$ is written down as follows
\begin{equation}
\left( \frac{du}{d\phi} \right)^2 = P_3(u) = b_3 u^3 + b_2 u^2 + b_1 u + b_0  
\end{equation}
where the corresponding coefficients are computed to be
\begin{align}
b_3 &=2M \\
b_2 &= -(1+\eta) \\
b_1 &= -\gamma\\
b_0 &=\frac{1}{3}\Lambda + \left( \frac{E}{L}\right)^2 \\
\end{align}
In order to obtain the solution in terms of the Weierstra{\ss} function, $\wp(\phi-\phi_{in};g_2,g_3)$,
first we perform a linear transformation of the form $u(\phi)=A y(\phi) + B$, where the coefficients
$\{A,B\}$ are given by
\begin{eqnarray}
A & = & \frac{4}{b_3} \\
B & = & - \frac{b_2}{3 b_3}
\end{eqnarray}
The initial equation $(du/d\phi)^2=P_3(u)$ takes the form \cite{thesis}
\begin{equation}
\left( \frac{dy}{d\phi} \right)^2 = 4y^3-g_2 y-g_3
\end{equation}
where the Weierstra{\ss} cubic invariants $g_2,g_3$ can be found \cite{thesis}
\begin{eqnarray}
g_2 & = & \frac{1}{16} \left( \frac{4b_2^2}{3} - 4 b_1 b_3  \right) \\
g_3 & = & \frac{1}{16} \left( \frac{b_1 b_2 b_3}{3} - b_0 b_3^2 - \frac{2 b_2^3}{27}  \right)
\end{eqnarray}
To obtain a complete solution of the Weierstra{\ss} function, $\wp(\phi-\phi_{in};g_2,g_3)$, the integration constant, $\phi_{in}$, must be determined. The last can be made by imposing the initial condition $u(\phi_0)=u_0$.
The solution to the initial equation is then given by
\begin{equation} \label{main1}
%\boxed{
r(\phi) =\frac{1}{A y(\phi) + B}=\frac{1}{A \: \wp(\phi-\phi_{in};g_2,g_3) + B}
%}
\end{equation}
provided that the discriminant 
\begin{equation}
\Delta \equiv g_2^3 - 27 g_3^2
\end{equation}
does not vanish, since when $\Delta=0$ the case is singular \cite{Kraniotis:2003ig}. For more details on the Weierstra{\ss} function and its properties, the interested reader may want to consult \cite{lectures}. For those who are more familiar with the Jacobi elliptic functions, the Weierstra{\ss} function may be expressed in terms of them, see e.g. \cite{elliptic}.

\smallskip

We recall that the formalism used up to now is valid for a family of geometries, which are
static, spherically symmetric solutions in massive gravity, in Weyl conformal gravity and in scale-dependent gravity (Schwarzschild-de Sitter with varying cosmological constant and Newton's constant). To investigate those cases, we will 
show below in a series of figures both the effective potential for photons as well as the corresponding orbits for each case, varying the energy, the initial angle and the running parameter $\xi$. 

\smallskip

\begin{figure*}[ht!]
\centering
\includegraphics[width=0.32\textwidth]{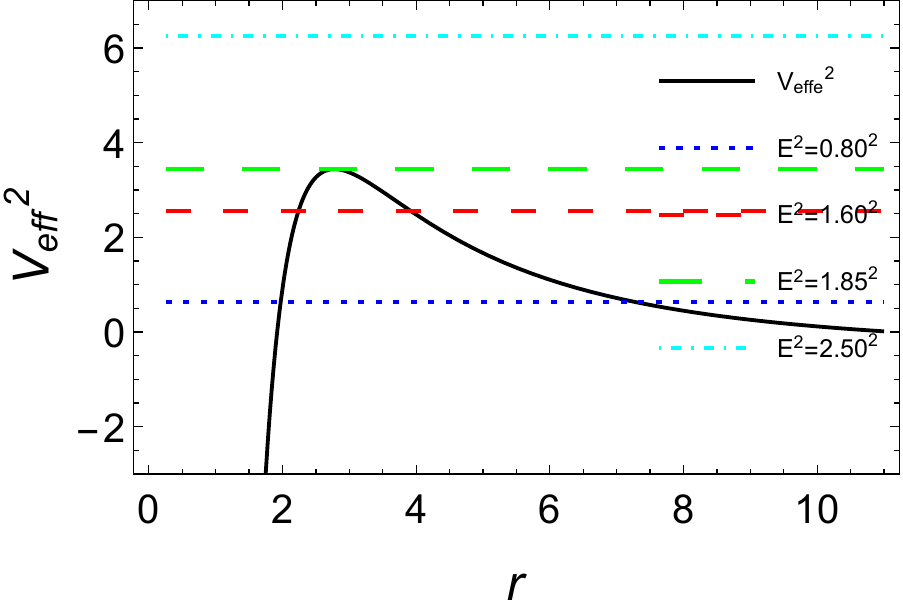}   \ \
%\\
\includegraphics[width=0.32\textwidth]{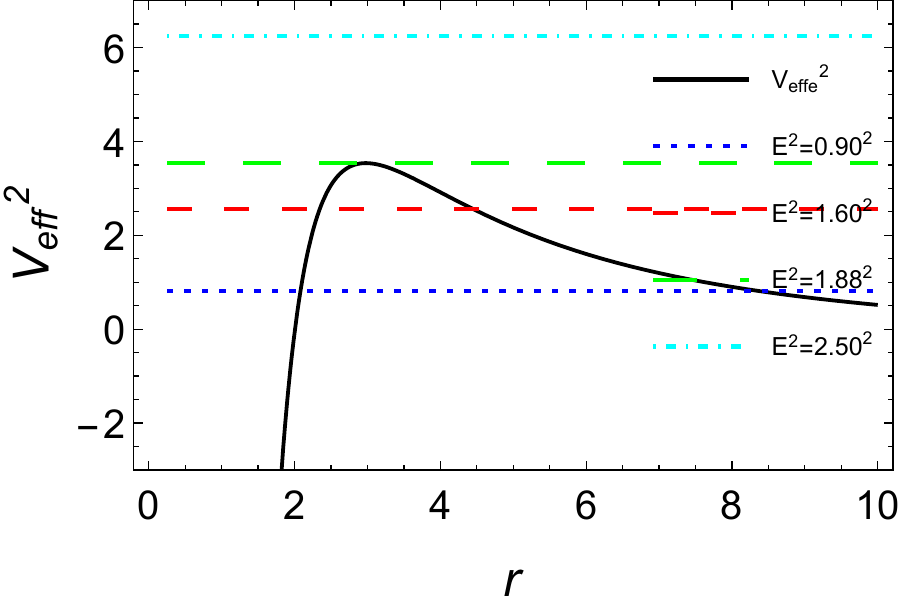}   \ \
\includegraphics[width=0.32\textwidth]{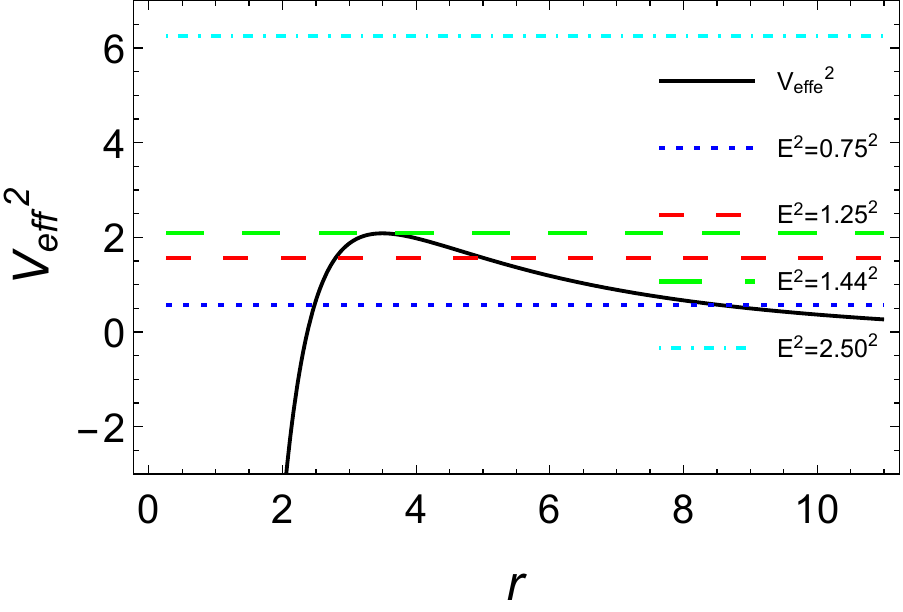}  
%}
\caption{ 
Effective potential for photons in the case of the class of geometries including three space-times considered in the first case setting $L=10, M=1, \Lambda=0.01$.
We plot, in each panel, the effective potential for photons and different energy regimes:
%i) {\textbf{Left Panel:}} Schwarzschild-de Sitter solution
i) {\textbf{Left Panel:}} Scale-dependent Schwarzschild-de Sitter setting $\xi=0.05$.
ii) {\textbf{Middle Panel:}} Massive gravity setting $\gamma=0.005, \eta=0$.
iii) {\textbf{Right Panel:}} Weyl conformal gravity setting $\gamma=0.005, \eta=-0.15$.
}
\label{fig:potentialscase1}
\end{figure*}

\begin{figure*}[ht!]
\centering
\includegraphics[width=0.48\textwidth]{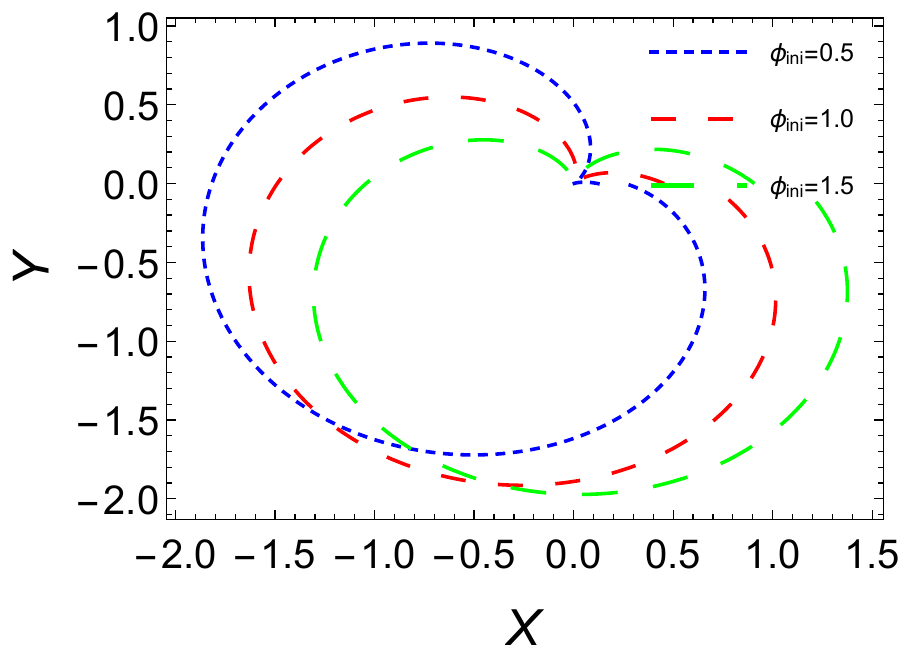}   \ \
\includegraphics[width=0.48\textwidth]{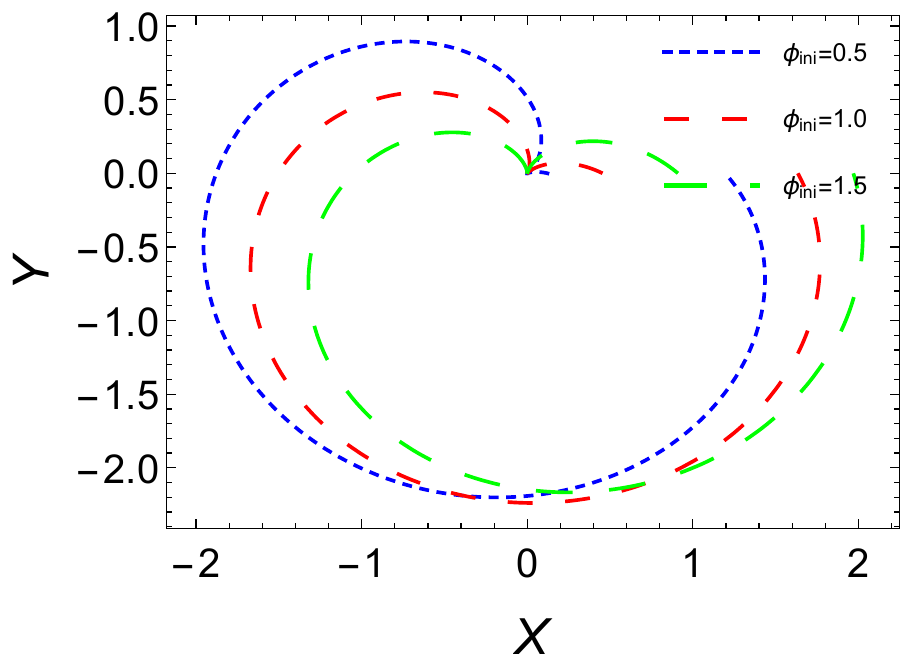}  
\\
\includegraphics[width=0.48\textwidth]{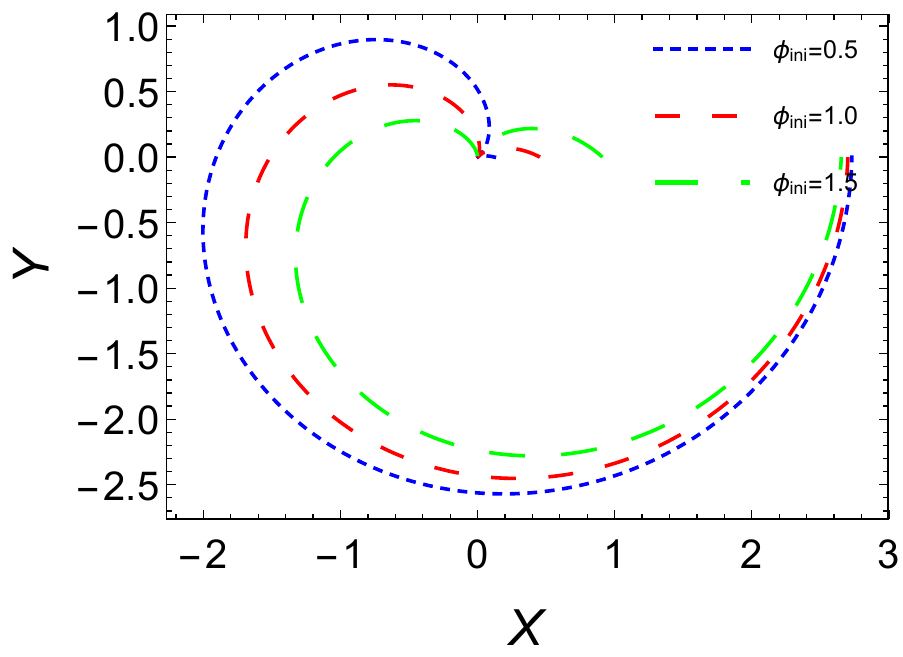}   \ \
\includegraphics[width=0.48\textwidth]{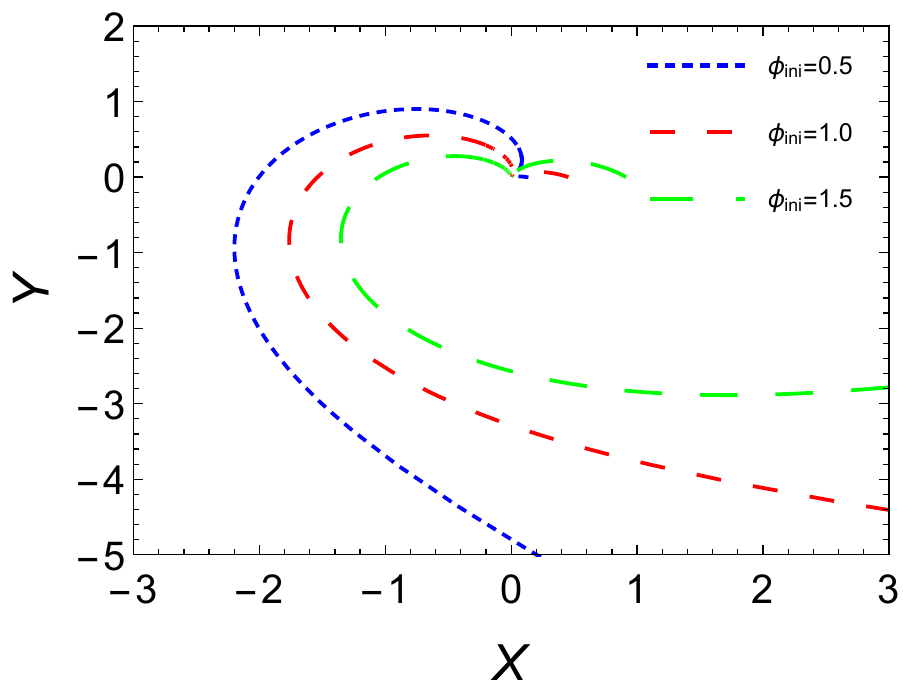}  
%}
\caption{ 
Photon orbits in the $(x-y)$ plane ($0 \leq \phi \leq 2 \pi$) in the case of scale-dependent Schwarzschild-de Sitter geometry setting $L=10, M=1, \Lambda=0.01, \xi=0.05$.
To show the impact of the photon energy and initial conditions, we take four different energies, and vary the initial angle as follows:
i) {\textbf{Top-left Panel:}} $E=0.80$ and $\phi_{ini} = \{0.5, 1.0, 1.5\}$ (short dashing blue line, dashing red line and long dashing green line, respectively).
ii) {\textbf{Top-right Panel:}} $E=1.60$ and $\phi_{ini} = \{0.5, 1.0, 1.5\}$ (short dashing blue line, dashing red line and long dashing green line, respectively).
iii) {\textbf{Bottom-left Panel:}} $E=1.85$ and $\phi_{ini} = \{0.5, 1.0, 1.5\}$ (short dashing blue line, dashing red line and long dashing green line, respectively).
iv) {\textbf{Bottom-right Panel:}} $E=2.50$ and $\phi_{ini} = \{0.5, 1.0, 1.5\}$ (short dashing blue line, dashing red line and long dashing green line, respectively).
}
\label{fig:1}
\end{figure*}

\begin{figure*}[ht!]
\centering
\includegraphics[width=0.48\textwidth]{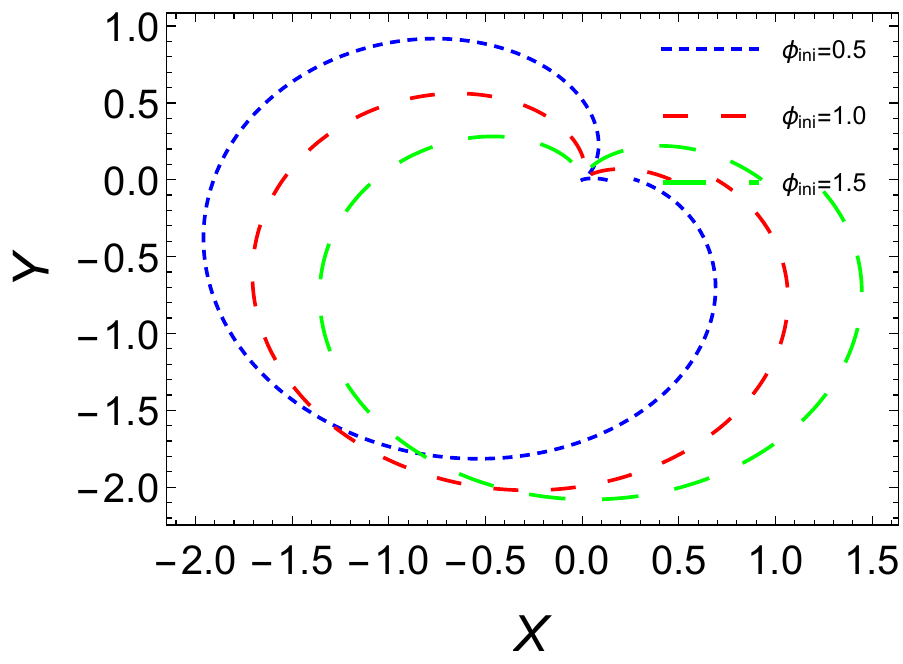}   \ \
\includegraphics[width=0.48\textwidth]{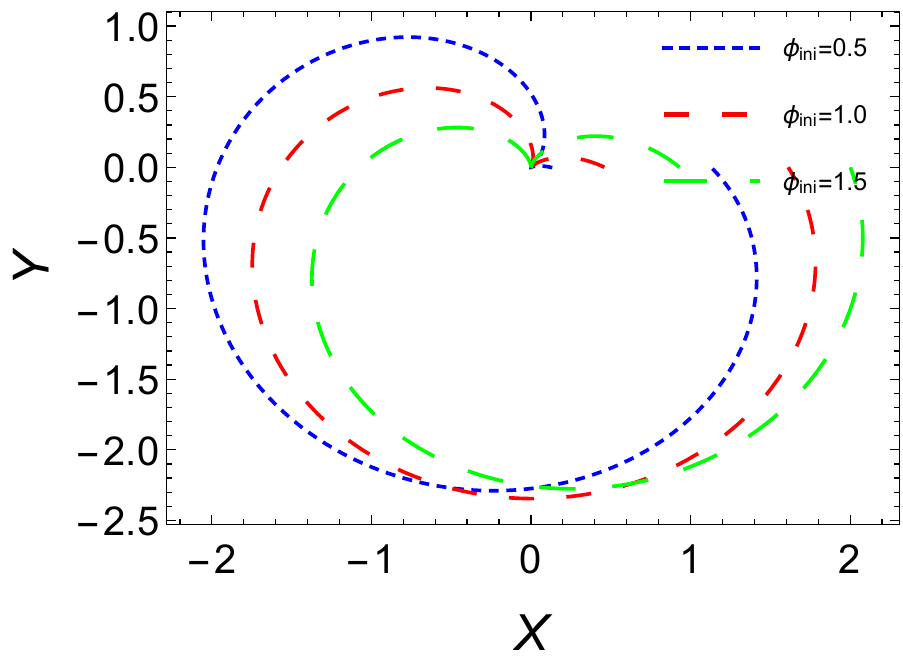}  
\\
\includegraphics[width=0.48\textwidth]{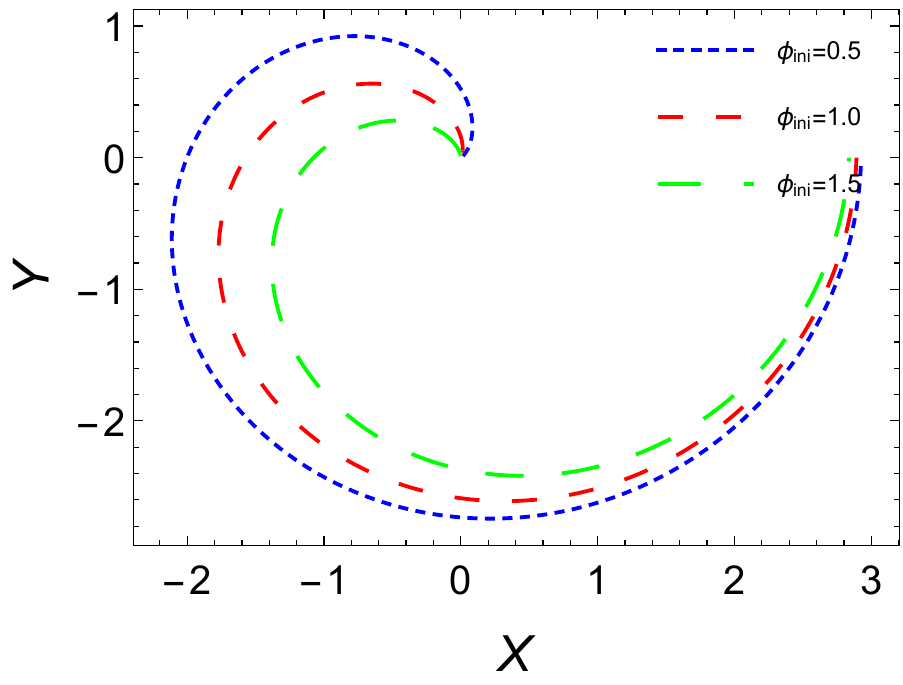}   \ \
\includegraphics[width=0.48\textwidth]{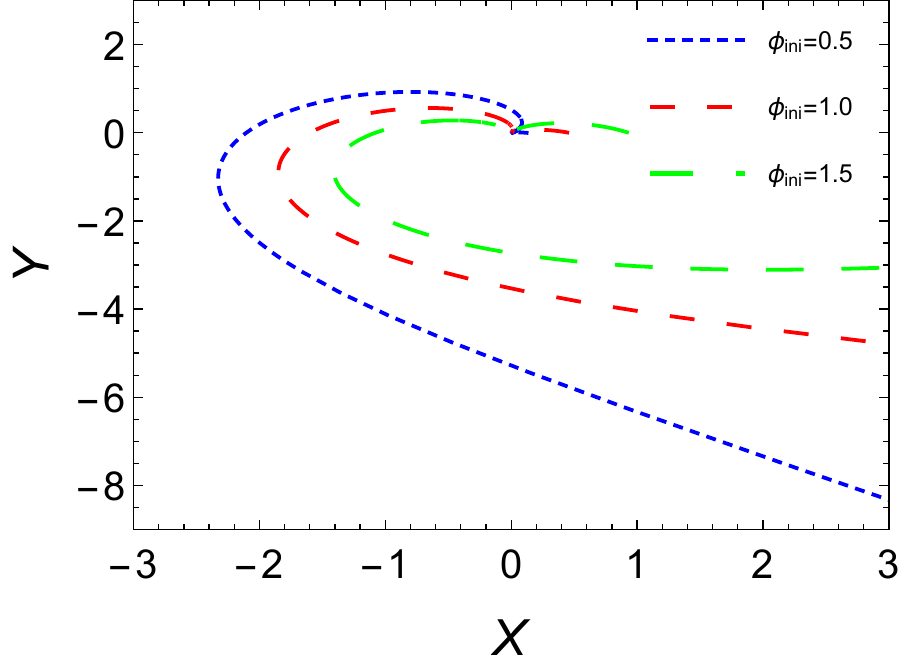}  
%}
\caption{ 
Photon orbits in the $(x-y)$ plane ($0 \leq \phi \leq 2 \pi$) in the case of massive gravity setting $L=10, M=1, \Lambda=0.01, \gamma=0.005, \eta = 0$.
To show the impact of the photon energy and initial conditions, we consider four different energies, and vary the initial angle as follows:
i) {\textbf{Top-left Panel:}} $E=0.90$ and $\phi_{ini} = \{0.5, 1.0, 1.5\}$ (short dashing blue line, dashing red line and long dashing green line, respectively).
ii) {\textbf{Top-right Panel:}} $E=1.60$ and $\phi_{ini} = \{0.5, 1.0, 1.5\}$ (short dashing blue line, dashing red line and long dashing green line, respectively).
iii) {\textbf{Bottom-left Panel:}} $E=1.88$ and $\phi_{ini} = \{0.5, 1.0, 1.5\}$ (short dashing blue line, dashing red line and long dashing green line, respectively).
iv) {\textbf{Bottom-right Panel:}} $E=2.50$ and $\phi_{ini} = \{0.5, 1.0, 1.5\}$ (short dashing blue line, dashing red line and long dashing green line, respectively).
}
\label{fig:Massive}
\end{figure*}

\begin{figure*}[ht!]
\centering
\includegraphics[width=0.48\textwidth]{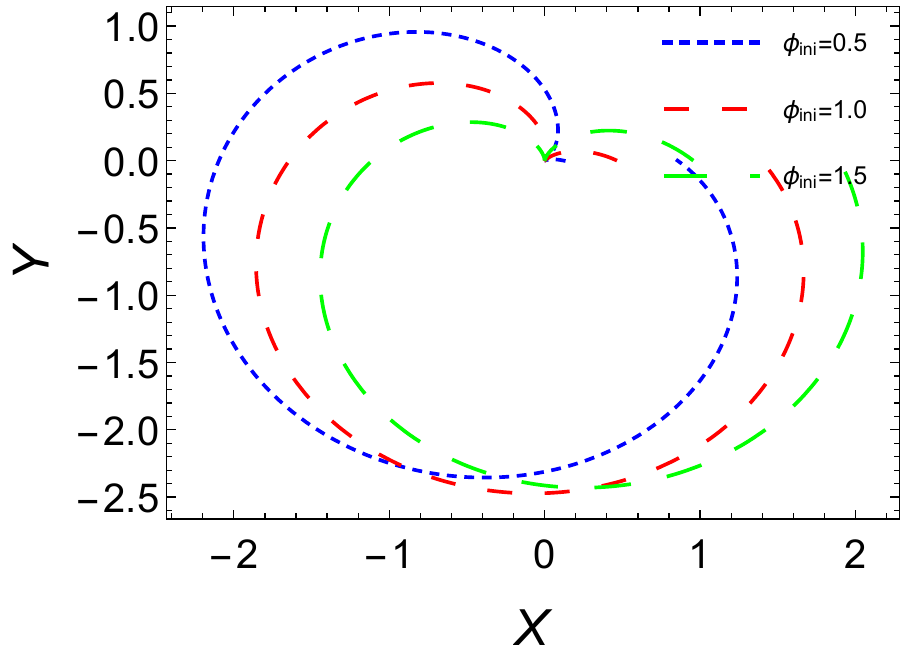}   \ \
\includegraphics[width=0.48\textwidth]{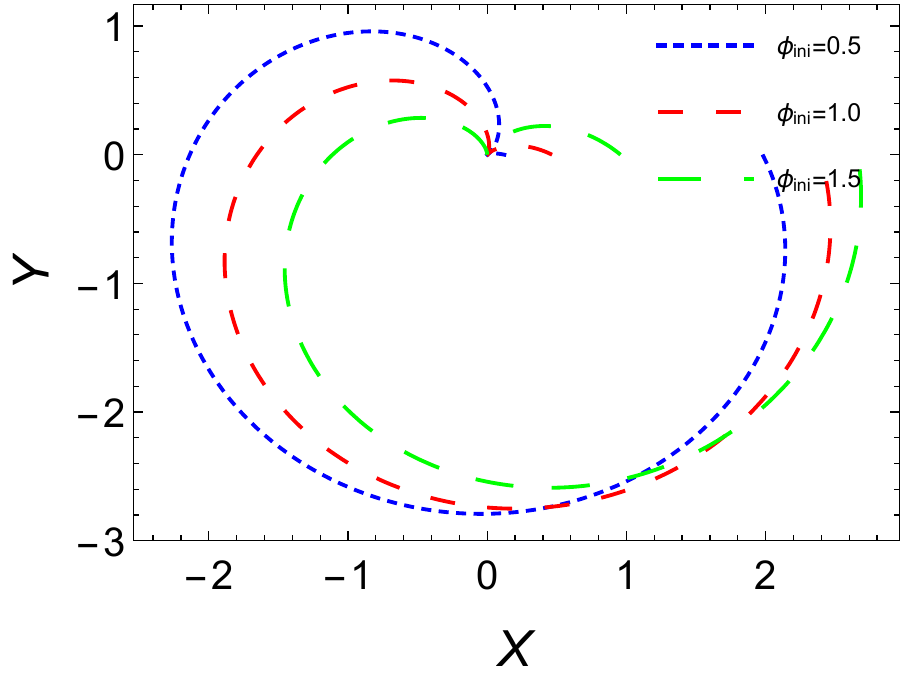}  
\\
\includegraphics[width=0.48\textwidth]{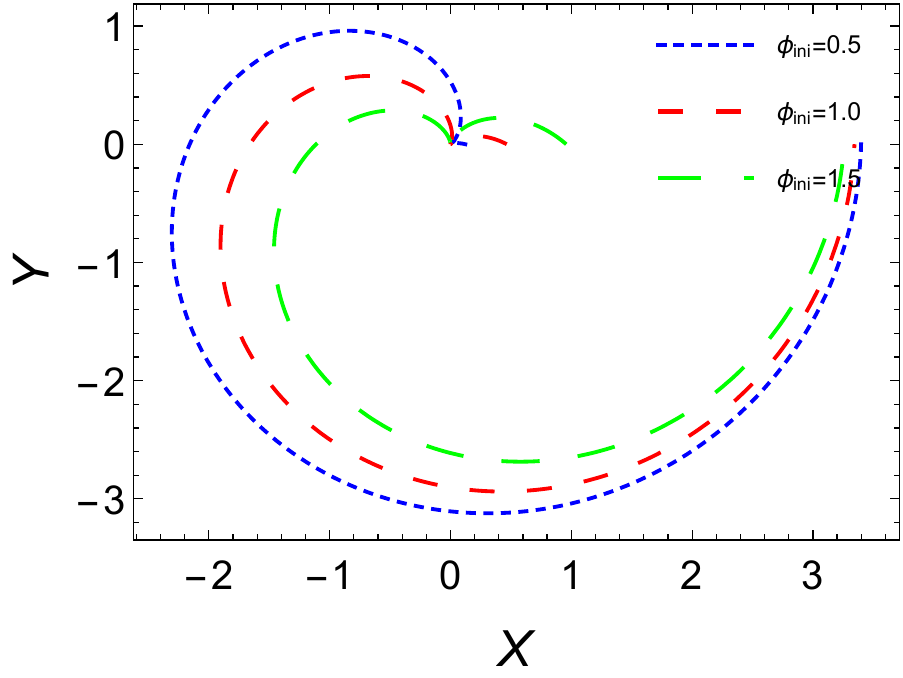}   \ \
\includegraphics[width=0.48\textwidth]{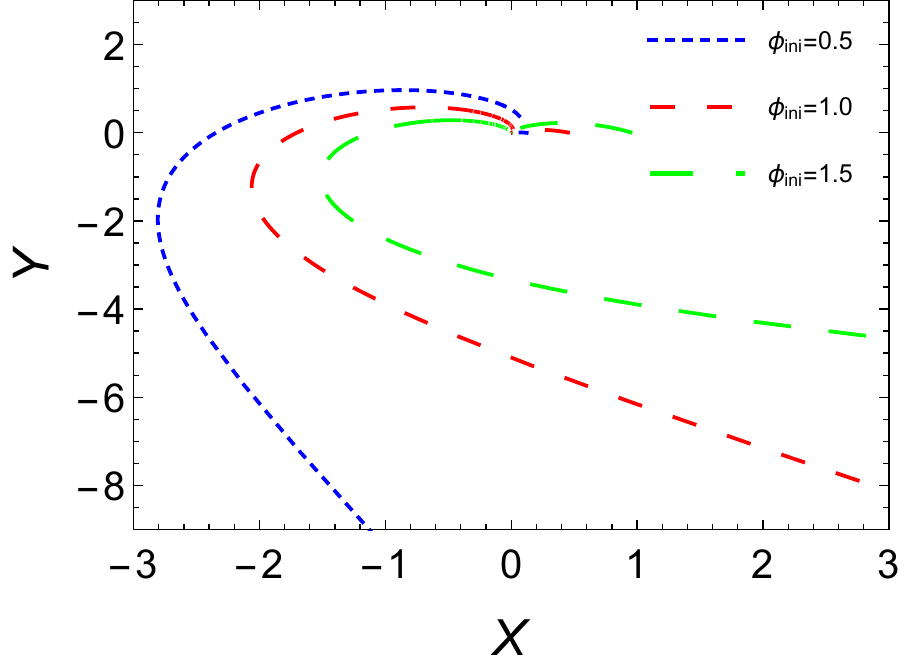}  
%}
\caption{ 
Photon orbits in the $(x-y)$ plane ($0 \leq \phi \leq 2 \pi$) in the case of Weyl gravity setting $L=10, M=1, \Lambda=0.01,
\gamma=0.005 \text{ and } \eta = -0.15$.
To show the impact of the energy and initial conditions, we take four different energies, and vary the initial angle as follows:
i) {\textbf{Top-left Panel:}} $E=0.75$ and $\phi_{ini} = \{0.5, 1.0, 1.5\}$ (short dashing blue line, dashing red line and long dashing green line, respectively).
ii) {\textbf{Top-right Panel:}} $E=1.25$ and $\phi_{ini} = \{0.5, 1.0, 1.5\}$ (short dashing blue line, dashing red line and long dashing green line, respectively).
iii) {\textbf{Bottom-left Panel:}} $E=1.44$ and $\phi_{ini} = \{0.5, 1.0, 1.5\}$ (short dashing blue line, dashing red line and long dashing green line, respectively).
iv) {\textbf{Bottom-right Panel:}} $E=2.50$ and $\phi_{ini} = \{0.5, 1.0, 1.5\}$ (short dashing blue line, dashing red line and long dashing green line, respectively).
}
\label{fig:Weyl}
\end{figure*}

In Fig. \eqref{fig:potentialscase1} we show the effective potential for each geometry, for different values of the energy. 
Subsequently, in Fig.~\eqref{fig:1} we show the trajectories for the scale-dependent Schwarzschild-de Sitter black hole. Besides, in figures \eqref{fig:Massive} and \eqref{fig:Weyl} we show the photon orbits in massive and Weyl gravity, varying the photon energy, $E$, and the initial angle $\phi_{in}$.

\smallskip

We have considered four different values of the energy, and for each one of those, we have assumed three different values of the initial angle. The orbits hit the (0,0) point and then bounce off, exhibiting a cardioid shape, which is deformed varying $E$ and $\phi_{in}$. The deformation becomes more pronounced as the photon energy approaches the maximum of the effective potential.

\smallskip

\begin{figure}[ht!]
\centering
\includegraphics[width=0.48\textwidth]{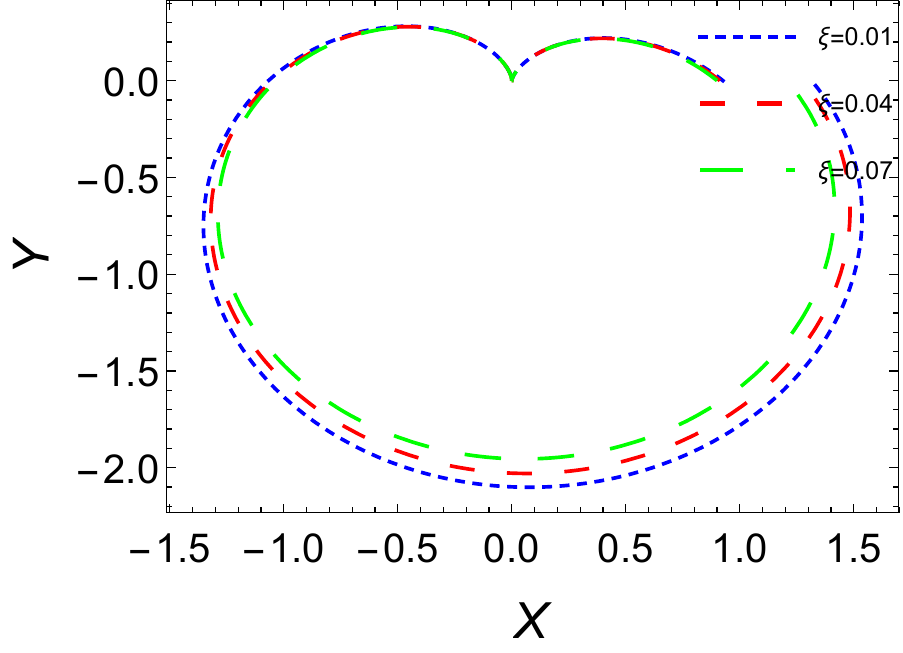}   
\caption{   
Photon orbits in the case of scale-dependent Schwarzschild-de Sitter geometry setting $L=10, M=1, \Lambda=0.01, E=1$ 
and $\xi=0.01,0.04,0.07$ from outwards to inwards.
}
\label{fig:ExtraPlot}
\end{figure}

Finally, in Fig.~\eqref{fig:ExtraPlot} we show the impact of the running parameter $\xi$ on the shape of the photon orbits (in the case of Schwarzschild-de Sitter geometry) setting $L=10, M=1, \Lambda=0.01, E=1$, and varying $\xi=0.01,0.04,0.07$ from outwards to inwards.

%%%%%%%%%%%%%%%%%%FIGURES%%%%%%%%%%%%%%%%%

%\begin{figure*}[ht!]
%\centering
%\includegraphics[width=0.48\textwidth]{trayeSTE090.pdf}   \ \
%\includegraphics[width=0.48\textwidth]{trayeSTE160.pdf}  
%\\
%\includegraphics[width=0.48\textwidth]{trayeSTE183.pdf}   \ \
%\includegraphics[width=0.48\textwidth]{trayeSTE250.pdf}  
%}
%\caption{ 
%Orbits of massless test particles in the $(x-y)$ plane ($0 \leq \phi \leq 2 \pi$) for the  Schwarzschild-de Sitter black hole with $L=10, M=1, \Lambda=0.01,
%\gamma=0 \text{ and } \eta = 0$.
%To show the impact of the energy and initial conditions, we take for different energies and vary the initial angle as follow:
%i) {\textbf{Top-left Panel:}} $E=0.90$ and $\phi_{ini} = \{0.5, 1.0, 1.5\}$ (short dashing blue line, dashing red line and long dashing green line, respectively).
%ii) {\textbf{Top-right Panel:}} $E=1.60$ and $\phi_{ini} = \{0.5, 1.0, 1.5\}$ (short dashing blue line, dashing red line and long dashing green line, respectively).
%iii) {\textbf{Bottom-left Panel:}} $E=1.84$ and $\phi_{ini} = \{0.5, 1.0, 1.5\}$ (short dashing blue line, dashing red line and long dashing green line, respectively).
%iv) {\textbf{Bottom-right Panel:}} $E=2.50$ and $\phi_{ini} = \{0.5, 1.0, 1.5\}$ (short dashing blue line, dashing red line and long dashing green line, respectively).
%%
%}
%\label{fig:SdS}
%\end{figure*}

%

%%%%%%%%%%%%%%%%%%%%%%%%%%%%%%%%%%%

%%%%%%%%%%%%%%%%%%%%%%%%%%%%%%%%%

%%%%%%%%%%%%%%%%%%%%%%%%%%%%

%%%%%%%%%%%%%%%%%%%%%%%%%%%%

\subsubsection{Case II}

\begin{figure}[ht!]
\centering
\includegraphics[width=0.48\textwidth]{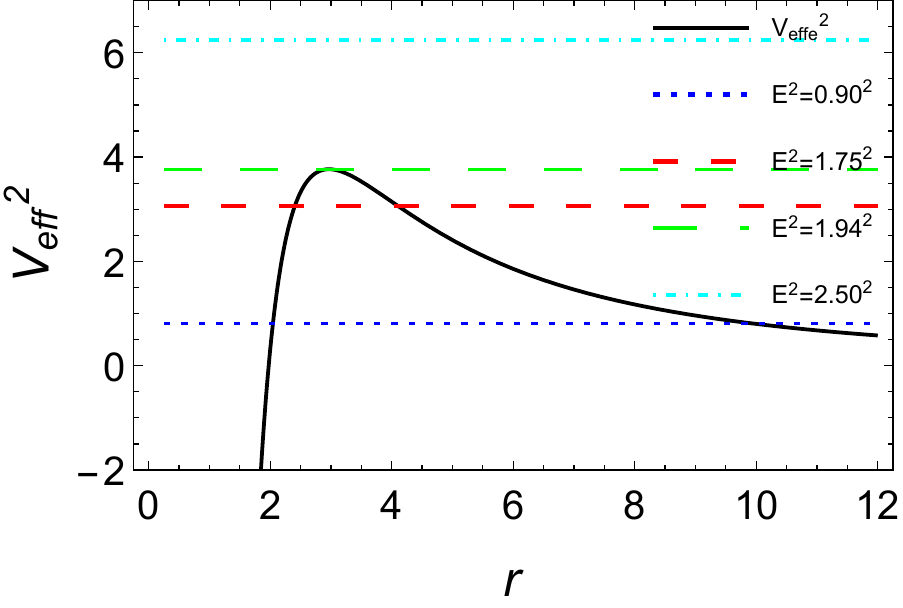}   
\caption{   
Effective potential for photons in the case of regular scale-dependent black holes setting $L=10, M=1, \xi=0.05$.
The different energy regimes are shown as well.
}
\label{fig:Potential_Regular}
\end{figure}

\begin{figure*}[ht!]
\centering
\includegraphics[width=0.48\textwidth]{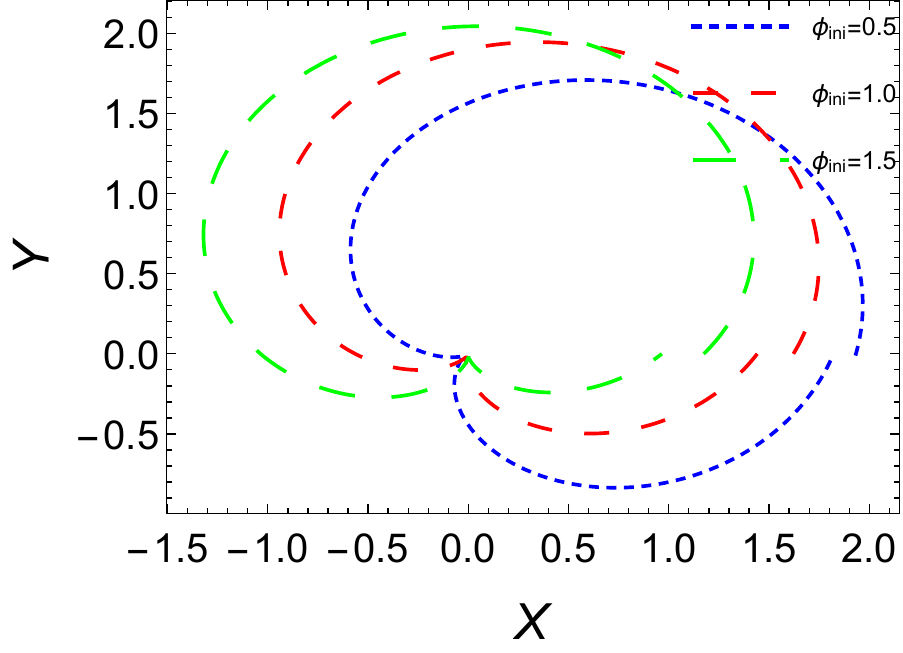}   \ \
\includegraphics[width=0.48\textwidth]{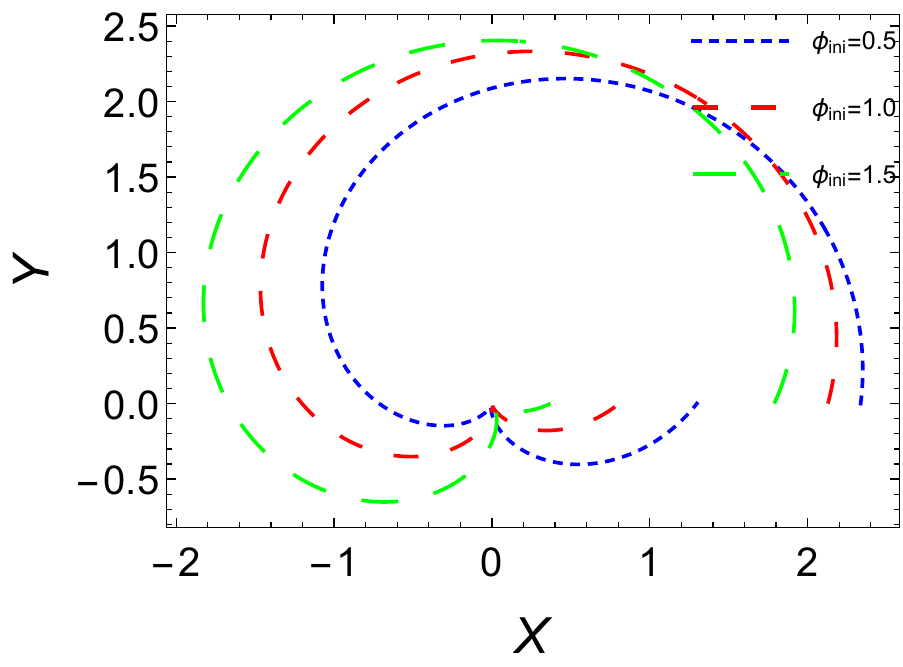}  
\\
\includegraphics[width=0.48\textwidth]{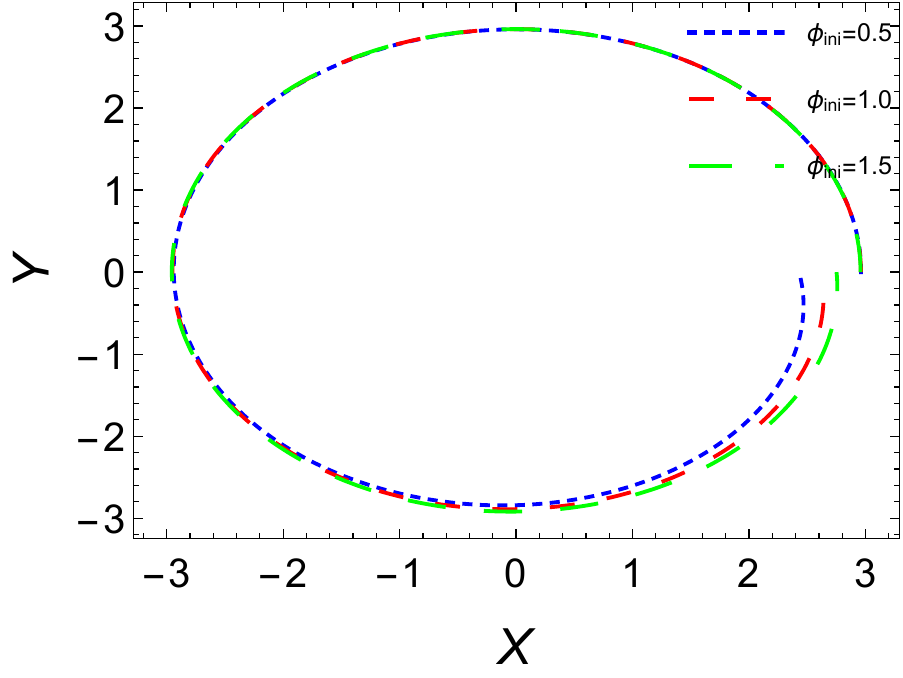}   \ \
\includegraphics[width=0.48\textwidth]{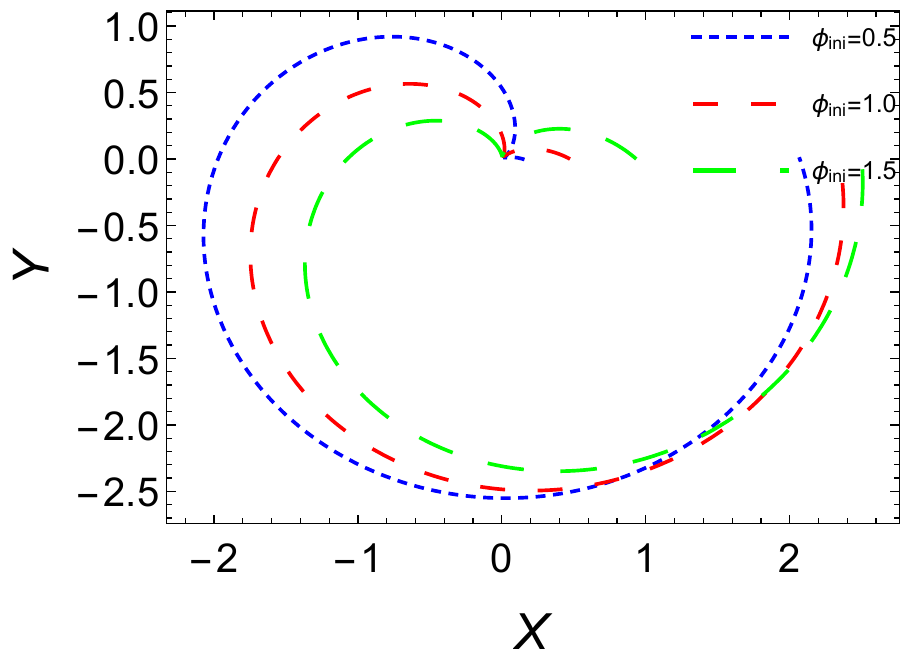}  
\caption{ 
Photon orbits in the $(x-y)$ plane ($0 \leq \phi \leq 2 \pi$) for a regular scale-dependent black hole with $L=10, M=1, \xi=0.05$
To show the impact of the energy and initial conditions, we take for different energies and vary the initial angle as follows:
i) {\textbf{Top-left Panel:}} $E=0.90$ and $\phi_{ini} = \{0.5, 1.0, 1.5\}$ (short dashing blue line, dashing red line and long dashing green line, respectively).
ii) {\textbf{Top-right Panel:}} $E=1.75$ and $\phi_{ini} = \{0.5, 1.0, 1.5\}$ (short dashing blue line, dashing red line and long dashing green line, respectively).
iii) {\textbf{Bottom-left Panel:}} $E=1.94$ and $\phi_{ini} = \{0.5, 1.0, 1.5\}$ (short dashing blue line, dashing red line and long dashing green line, respectively).
iv) {\textbf{Bottom-right Panel:}} $E=2.50$ and $\phi_{ini} = \{0.5, 1.0, 1.5\}$ (short dashing blue line, dashing red line and long dashing green line, respectively).
}
\label{fig:Regular}
\end{figure*}

In this case, the corresponding metric components are given by

\begin{align}
g_{tt} &=F(r) =  1 - \frac{2 M}{r} + \frac{M^2 \xi}{r^2}
\\
g_{rr} &= F(r)^{-1}
\end{align}
Similarly to the previous case, we can use the following change of variable $u=1/r$, and also setting $\epsilon=0$, 
the equation for the trajectories $u(\phi)$ is written down as follows 
\begin{equation}
\left( \frac{du}{d \phi} \right)^2 = P_4(u) = -M^2 \xi u^4 + 2 M u^3 - u^2 + (E/L)^2
\end{equation}
with $P_4(u)$ being a fourth degree polynomial in $u$. We can simplify the problem reducing the order of the polynomial 
if we take advantage of a single root of $P_4(u)$. Thus, we can set $u=(1/z)+l$ to get a new equation for $z(\phi)$ of the form \cite{thesis}
\begin{equation}
\left( \frac{dz}{d \phi} \right)^2 = b_3 z^3 + b_2 z^2 + b_1 z + b_0
\end{equation}
where now the coefficients are computed to be
\begin{align}
b_3 &=-4 l^3 M^2 \xi +6 l^2 M-2 l
\\
b_2 &= -6 l^2 M^2 \xi +6 l M-1
\\
b_1 &= 2 M-4 l M^2 \xi
\\
b_0 &= -M^2 \xi
\end{align}
The new equation may be now solved exactly as before, and therefore we find for $r(\phi)$ the expression
\begin{equation} \label{main2}
r(\phi)=\frac{A \: \wp(\phi-\phi_{in};g_2,g_3) + B}{1 + l \: [A \: \wp(\phi-\phi_{in};g_2,g_3) + B]}
\end{equation}
We then show in Fig. \eqref{fig:Potential_Regular} the effective potential for photons, and after that we plot 
the photon orbits for different values of photon energy and initial angle $\phi_{in}$, see Fig. \eqref{fig:Regular}).
Here, too, we have considered four different values of $E$, and three different values of $\phi_{in}$. All the features
observed in the geometries discussed before are observed here as well.

\smallskip

%%%%%%%%%%%%%%%%%%%FIGURES%%%%%%%%%%%%%%%%%%%%%%%%%%%%%%%%%

%%%%%%%%%%%%%%%%%%%%%%%%%%%%%%%%

%%%%%%%%%%%%%%%%%%%%%%%%%%%%%%%%%

\begin{figure}[ht!]
\centering
\includegraphics[width=0.48\textwidth]{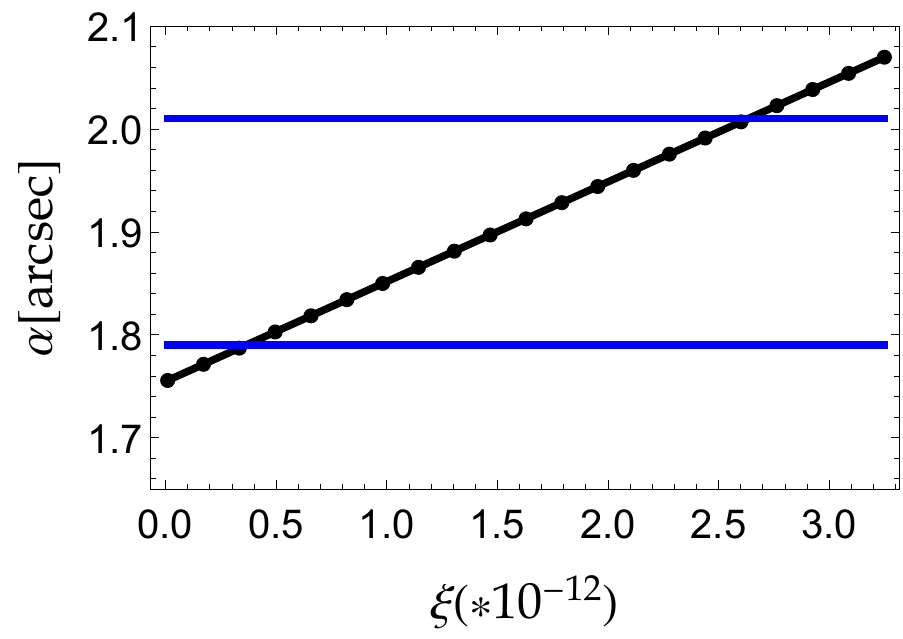}   
\caption{   
Prediction for light deflection near our Sun in the case of scale-dependent Schwarzschild-de Sitter geometry. 
We show the deflection angle (in arcsec) as a function of the running parameter $\xi$. The allowed strip $1.9 \pm 0.11$ is shown as well.
}
\label{fig:Deflection}
\end{figure}

%%%%%%%%%%%%%%%%%%%%%FIGURES%%%%%%%%%%%%%%%%%%%%%%%%%%%%%%%%%%%%%

%{\bf 
Before we conclude our work, let us briefly discuss some applications and observational consequences
based on the formalism used in the present article. In the strong field limit \cite{Bozza, Tsukamoto1} analytic expressions 
may be obtained, and observables related to gravitational lensing, such as image angle and magnification, are
computed in terms of the strong deflection coefficients, see e.g. \cite{Tsukamoto2,Sendra:2018vux}. For well-known
geometries, such as the Schwarzschild and Reissner-Nordstr{\"o}m space-times, the expressions for the strong deflection coefficients are shown in the Appendix of \cite{Torres}. A detailed investigation of the strong field limit for the geometries discussed here lies outside the scope of this work. It would be interesting, however, to
address that issue, and we certainly hope to be able to do so in a forthcoming publication. 

\smallskip

Finally, in the discussion to follow we shall use the data on the solar eclipse occurred in 1919, which was
measured by the Eddington expedition \cite{dataI}, and was reanalyzed much later in \cite{dataII}, to
constrain the running parameter $\xi$, which is the only free parameter in scale-dependent gravity as the 
cosmological constant as well as the solar mass are known. 

The deflection angle is given by \cite{Tsukamoto2}
\begin{equation}
\alpha = -\pi + 2 \int_{r_0}^\infty \: \frac{dr}{\sqrt{P_3(r)}}
\end{equation}
where $r_0$ is the closest distance to the deflecting object, determined by $dr/d \phi = 0$, or equivalently, $P_3(r_0)=0$.
To compute the integral it is more convenient to introduce a new dimensionless parameter, $z \equiv 1-r_0/r$. 
The deflection angle now is given by
\begin{equation}
\alpha = -\pi + \int_{0}^1 \: \frac{2 r_0 dz}{\sqrt{c_1 z+c_2 z^2+c_3 z^3}}
\end{equation}
where the coefficients $c_i$ are found to be
\begin{eqnarray}
c_1 & = &  -6 M r_0 + 2 r_0^2 + \xi r_0^2 (6 r_0^2-r_0/M) \\
c_2 & = &  6 M r_0 - 2 r_0^2 - 3 \xi r_0^2  \\
c_3 & = & -2 M r_0      
\end{eqnarray}
The theoretical prediction for the deflection angle as a function of the running parameter, $\xi$, assuming an impact parameter 
\begin{equation}
b \equiv L/E = R_{\odot}=6.96 \times 10^8~m
\end{equation}
is shown in Fig.~\eqref{fig:Deflection}. The strip from 1.79 to 2.01 denotes the allowed observational range, $\alpha=1.9 \pm 0.11$ \cite{Crispino:2019yew}. Our results show that $\xi$ must not exceed the value $\xi_* = 2.6 \times 10^{-12}$.
Therefore we obtain for the first time here an upper bound for the running parameter
\begin{equation}
\xi \leq 2.6 \times 10^{-12}.
\end{equation}
%}

%%%%%%%%%%%%%%%%%%%%%%%
\section{Conclusion}
%%%%%%%%%%%%%%%%%%%%%%

We have studied the orbits of light rays in the gravitational background of $(1+3)$-dimensional 
geometries with spherical symmetry arising in theories other than General Relativity, such as 
scale-dependent gravity, massive gravity and Weyl conformal gravity. In particular, we have obtained 
exact analytical solutions to the null geodesic equations in terms of the Weierstra{\ss} function. 
The effective potential for photons as well as the trajectories in the $(x-y)$ plane have been shown graphically
for several values of the photon energy, the integration constant (initial conditions) and the running parameter. 
In the case of scale-dependent Schwarzschild-de Sitter geometry, the light deflection angle is computed as a function 
of the running parameter, and an upper bound for the latter is obtained.

\smallskip

This work opens to us the possibility of testing such classes of alternative theories of gravity at any scale, 
for instance, around neutron stars and astrophysical black holes, or around super-massive black holes at the center
of galaxies. More importantly, it may be used to generalize the gravitational lensing of galaxies, which is fundamental 
to further understand the content as well as other important aspects of the Universe.

%%%%%%%%%%%%%%%%%%%%%%%%%%
\section*{Acknowlegements}
%%%%%%%%%%%%%%%%%%%%%%%%%%%

We thank the anonymous reviewer for his/her suggestions. 
The authors G.~P. and I.~Lopes thank the Funda\c c\~ao para a Ci\^encia e Tecnologia (FCT), Portugal, 
for the financial support to the Center for Astrophysics and Gravitation-CENTRA, Instituto Superior T\'ecnico, 
Universidade de Lisboa, through the Project No.~UIDB/00099/2020 and grant No. PTDC/FIS-AST/28920/2017.
The author A.~R. acknowledges Universidad de Tarapac\'a for financial support.

%%%%%%%%%%%%%%%%%%%%%%%%%%%%%%%%%%%%%%%%%%%%%%%%%%%%%%%%%%%%%%%%%%%%%%%%%%%%%%%%%%%%%%%%%%%%%%

\end{document}